%% file: main/main.tex
\documentclass[sigconf]{acmart} 
\AtBeginDocument{%
  }

\copyrightyear{2025}
\acmYear{2025}
\acmConference[MIND '25]{Next-Gen Middleware for MLOps in Distributed
Systems }{December 15--19, 2025}{Nashville, TN, USA}
\input{./main/config.tex}

\usepackage{microtype}

\begin{document}

\title[Automated Dynamic AI Inference Scaling on HPC-Infrastructure]{Automated Dynamic AI Inference Scaling on HPC-Infrastructure: Integrating Kubernetes, Slurm and vLLM}



\author{Tim Trappen}
\affiliation{%
  \institution{Ruhr University Bochum}
  \city{Bochum}
  \country{Germany}
}
\email{tim.trappen@ruhr-uni-bochum.de}

\author{Robert Keßler}
\affiliation{%
  \institution{University of Cologne}
  \city{Cologne}
  \country{Germany}}
\email{kessler@uni-koeln.de}

\author{Roland Pabel}
\affiliation{%
  \institution{University of Cologne}
  \city{Cologne}
  \country{Germany}}
\email{pabel@uni-koeln.de}

\author{Viktor Achter}
\affiliation{%
  \institution{University of Cologne}
  \city{Cologne}
  \country{Germany}}
\email{achter@uni-koeln.de}

\author{Stefan Wesner}
\affiliation{%
  \institution{University of Cologne}
  \city{Cologne}
  \country{Germany}}
\email{wesner@uni-koeln.de}





\renewcommand{\shortauthors}{Trappen et al.}

\begin{abstract}
  \input{./content/abstract.tex}
\end{abstract}



\maketitle

\input{./content/01_introduction.tex}
\input{./content/02_related_work.tex}

\input{./content/03_architecture.tex}
\input{./content/04_preliminary_results.tex}
\input{./content/04_discussion.tex}
\input{./content/05_conclusion.tex}
\input{./content/acknowledgements.tex}

\bibliographystyle{ACM-Reference-Format}
\bibliography{./main/bibliography}

\end{document}

%% file: main/config.tex
\usepackage{times}
\usepackage{latexsym}

\usepackage{graphicx}

\usepackage{hyperref}

\usepackage{booktabs}
\usepackage{multirow}

\usepackage[nolist]{acronym}
\input{./content/acronyms}
\usepackage{cleveref}

\usepackage{svg}
\usepackage{amsmath}

%% file: content/acronyms.tex
\begin{acronym}
    \acro{AI}{Artificial Intelligence}
    \acro{BT}{Block Tridiagonal}
    \acro{BWA}{Burrows-Wheeler Aligner}
    \acro{CCD}{Core Chiplet Die}
    \acro{CCX}{Core Complex}
    \acro{CG}{Conjugate Gradient}
    \acro{CPU}{Central Processing Unit}
    \acro{CVM}{Confidential Virtual Machines}
    \acro{DBMS}{Database Management System}
    \acro{DNA}{Deoxyribonucleic Acid}
    \acro{DPU}{Data Processing Unit}
    \acro{DRAM}{Dynamic Random-Access Memory}
    \acro{EP}{Embarrassingly Parallel}
    \acro{EPC}{Enclave Page Cache}
    \acro{EU}{European Union}
    \acro{FLOPS}{Floating Point Operations Per Second}
    \acro{FT}{Fast Fourier Transform}
    \acro{GPFS}{General Parallel File System}
    \acro{GPU}{Graphics Processing Unit}
    \acro{HPC}{High-Performance Computing}
    \acro{IoT}{Internet of Things}
    \acro{IS}{Integer Sort}
    \acro{JVM}{Java Virtual Machine}
    \acro{KVC}{Key-Value Cache}
    \acro{KVM}{Kernel-based Virtual Machine}
    \acro{LLM}{Large Language Model}
    \acro{LU}{LU Decomposition}
    \acro{ML}{Machine Learning}
    \acro{MG}{MultiGrid}
    \acro{NGS}{Next-Generation Sequencing}
    \acro{NPB}{NAS Parallel Benchmarks}
    \acro{NUMA}{Non-Uniform Memory Access}
    \acro{OS}{Operating System}
    \acro{RNA}{Ribonucleic Acid}
    \acro{SEV}{Secure Encrypted Virtualization}
    \acro{SEV-ES}{Secure Encrypted Virtualization–Encrypted State}
    \acro{SEV-SNP}{Secure Encrypted Virtualization–Secure Nested Paging}
    \acro{SGX}{Software Guard Extensions}
    \acro{SME}{Secure Memory Encryption}
    \acro{SMP}{Symmetric Multiprocessing}
    \acro{SNP}{Secure Nested Paging}
    \acro{SP}{Scalar Pentadiagonal}
    \acro{TCB}{Trusted Computing Base}
    \acro{TD}{Trust Domain}
    \acro{TDX}{Trust Domain Extensions}
    \acro{TEE}{Trusted Execution Environment}
    \acro{TPU}{Tensor Processing Unit}
    \acro{UA}{Unstructured Adaptive}
    \acro{VM}{Virtual Machine}
    \acro{VPN}{Virtual Private Network}
\end{acronym}

%% file: content/abstract.tex
Due to rising demands for Artificial Inteligence (AI) inference, especially in higher education, novel solutions utilising existing infrastructure are emerging. The utilisation of High-Performance Computing (HPC) has become a prevalent approach for the implementation of such solutions. However, the classical operating model of HPC does not adapt well to the requirements of synchronous, user-facing dynamic AI application workloads. In this paper, we propose our solution that serves LLMs by integrating vLLM, Slurm and Kubernetes on the supercomputer \textit{RAMSES}. The initial benchmark indicates that the proposed architecture scales efficiently for 100, 500 and 1000 concurrent requests, incurring only an overhead of approximately 500 ms in terms of end-to-end latency. 


%% file: content/01_introduction.tex
\section{Introduction}
\ac{AI}, and in particular \acp{LLM}, have become an integral factor in numerous areas of our lives. 
For instance, the use of \acp{LLM} in an university setting involves a wide range of user groups, including students, researchers, and even the administration. Students benefit from  personalized education pathways, are provided with dynamic learning content and feedback, and thus can self-regulate their individual progress \cite{bernalRevolutionizingELearningAssessments2024, liuOptimizingELearningEnvironments2024}. The use of chatbots within e-learning platforms, such as MoodleBot, provides students with a personal tutor who is available 24/7, thereby reducing the administrative workload for teachers \cite{neumannLLMDrivenChatbotHigher2025, liuTeachingCS50AI2024}. Scientists are already making use of the increasingly powerful \acp{LLM} for their research work in various domains such as drug discovery or partial differential equations, where the requests range from simple literature search and summary to concept clarification, data analysis and methodology guidance \cite{ai4scienceImpactLargeLanguage2023}. However, the potential of \acp{LLM} goes way beyond this, meaning that in the future they could serve as a kind of \ac{AI} research assistant, as envisaged by the AuroraGPT \footnote{\url{https://www.anl.gov/cels/auroragpt-foundation-models-for-science}} project \cite{, cappelloAuroraGPTExploringAI2024}. Even the administrative offices are increasingly using AI tools, for example to support the enrollment process or to provide better assistance with scholarship applications \cite{khairullahImplementingArtificialIntelligence2025}.
\par For higher education institutions, compliance and data governance pose the greatest challenges when adopting to \ac{AI}. This creates the need for independent and sovereign \ac{AI} infrastructure which provides easy to use and privacy preservering access to models. Universities that operate their own data centers are predestined to offer such services themselves, as they tend to possess the necessary computing power and can also guarantee the data privacy aspects. The computing systems operated by higher education institutions are often \ac{HPC} systems, also known as supercomputers, which are used by scientists from a wide range of scientific disciplines to solve highly complex research problems and perform simulations, but tend to be underutilized \cite{10740905}. However, the operating model of \ac{HPC} systems is based on static allocation of computing resources, which is suitable for training \acp{LLM} but is not an inherent match for processing dynamic \ac{LLM} inference requests \cite{lopezAIFactoriesIts2025}. To overcome this discrepancy, one can either \textit{(i)} implement dynamic resource allocation in \ac{HPC} systems from scratch, which requires at least adjustments at the level of the resource manager and programming models \cite{tarrafMalleabilityModernHPC2024}, or \textit{(ii)} implement a scalable web API, that handles the dynamic management of HPC jobs and forwards the inference requests accordingly. Due to the manifold modifications required for approach \textit{(i)}, this comprehensive implementation is very sophisticated, whereas approach \textit{(ii)} offers an implementation using micro-service architecture, which can be deployed in a scalable manner using e.g. Kubernetes.

\par The remainder of this paper is structured as follows, in \cref{sec:related_work} we present and discuss challenges and other approaches to solve the problem of serving dynamic \ac{AI} inference workloads on \ac{HPC} infrastructure. In \cref{sec:architecture} we describe our own approach, evaluate its performance in \cref{sec:evaluation} and discuss eventual threats to validity in \cref{sec:discussion}. We conclude our paper with an outlook on the future work in \cref{sec:outlook}.

%% file: content/02_related_work.tex
\section{Related Work}
\label{sec:related_work}
As \ac{HPC} and \ac{AI} converge, this is reflected, for example, in the  massive investments of the \ac{EU} in so-called \textit{AI Factories}\footnote{\url{https://digital-strategy.ec.europa.eu/en/policies/ai-factories}}, which are intended to drive forward the development of trustworthy generative \ac{AI} models in Europe. But also on a smaller scale, existing \ac{HPC} infrastructures at higher education institutions are being used to run \ac{AI} projects such as \textit{Open Source-KI.nrw} \footnote{\url{https://www.oski.nrw}}. 
This poses a certain contradiction, as \ac{AI} environments are usually based on dynamic cloud-native solutions, whereas \ac{HPC} systems are operated using batch processing and static resource allocation. 
Specifically, \ac{HPC} systems are not inherently designed to support interactive applications, which is why \citeauthor{lopezAIFactoriesIts2025} propose to rely on a sovereign dual-stack architecture for future systems, in order to combine the best of both worlds \cite{lopezAIFactoriesIts2025}.
\ac{HPC} systems have always been performance-sensitive, yet with the advent of AI workloads, it is becoming critical for cloud computing as well. In order to ensure sustainable performance of \ac{HPC} systems, especially in the exascale era, accelerators such as \acp{GPU} are essential, which in turn are also required for \ac{AI}. The approach proposed by \citeauthor{hoeflerXaaSAccelerationService2024}, aims on the advatages of the \ac{HPC}-Cloud convergence, by providing a simple and fast access to accelerators with their Acceleration as a Service (XaaS) model \cite{hoeflerXaaSAccelerationService2024}. In turn, AI itself can be used in the HPC domain to efficiently generate high-performant parallel code, which subsequently improves not only conventional HPC applications but also the models themselves \cite{chenLandscapeChallengesHPC2024}.
\par This kind of convergence does also pose certain challenges, including not only performance but also sustainability and scalability factors. \citeauthor{reschFutureHPCAI2025} therefore suggests that, in addition to the traditional TOP500 list of supercomputers, future evaluations should also focus specifically on \ac{AI} workloads, emphasizing the time to deliver a solution or the complexity of finding a solution rather than pure \ac{FLOPS} \cite{reschFutureHPCAI2025}. In order to overcome the challenges associated with realizing \ac{AI}-coupled \ac{HPC} workflows, the development of new middleware solutions is required to accommodate various execution motifs such as dynamic orchestration, multistage pipeline, and adaptive training \cite{brewerAIcoupledHPCWorkflow2025}. 

To provide \ac{AI} inference on a system, a variety of services must be operated and linked with each other, e.g., a user registry, service registry, and service identifier components, but also the support of scheduling and scaling functions. \citeauthor{guranMiddlewareLargeLanguage2024} present a middleware solution that considers this and serves as a \ac{LLM} gateway, but is lacking support for \ac{HPC} environments \cite{guranMiddlewareLargeLanguage2024}. 
The \textit{FLEXI} project at FernUniversiät in Hagen implements on-premise hosting of open-source \acp{LLM} in an higher-education environment but also lacks support for \ac{HPC} systems \cite{zeschFlexibleLLMExperimental2024}.
This shortcoming has been addressed by \citeauthor{luizScalableEnginePerformance2025}, who have developed a solution that implements a scalable microservice-based master-slave architecture that serves
heterogeneous \ac{LLM} models on \ac{HPC} infrastructure leveraging the Slurm workload manager \cite{luizScalableEnginePerformance2025}. A similar implementation is offered by \textit{Chat AI}, which operates its chat front-end as a web service in the Cloud , and then forwards the corresponding \ac{LLM} requests via an API gateway to the respective model on the \ac{HPC} infrastructure for being processed 
\cite{doosthosseiniChatAISeamless2024a}.
\par Despite all these efforts, resource and workflow managers still show a lack of support for the openness, transparency, and reusability of such workflows on HPC systems, meaning that the inherent dynamicity in such systems needs to be further fostered \cite{ejarqueEnablingDynamicIntelligent2022b}. 
Furthermore, although attempts have been made to implement AI inference on HPC systems, there is a lack of systematic performance studies \cite{brewerInferenceBenchmarkingHPC2020}.

%% file: content/03_architecture.tex
\section{Architecture}
\label{sec:architecture}
\begin{figure}[ht]
  \includegraphics[width=\columnwidth]{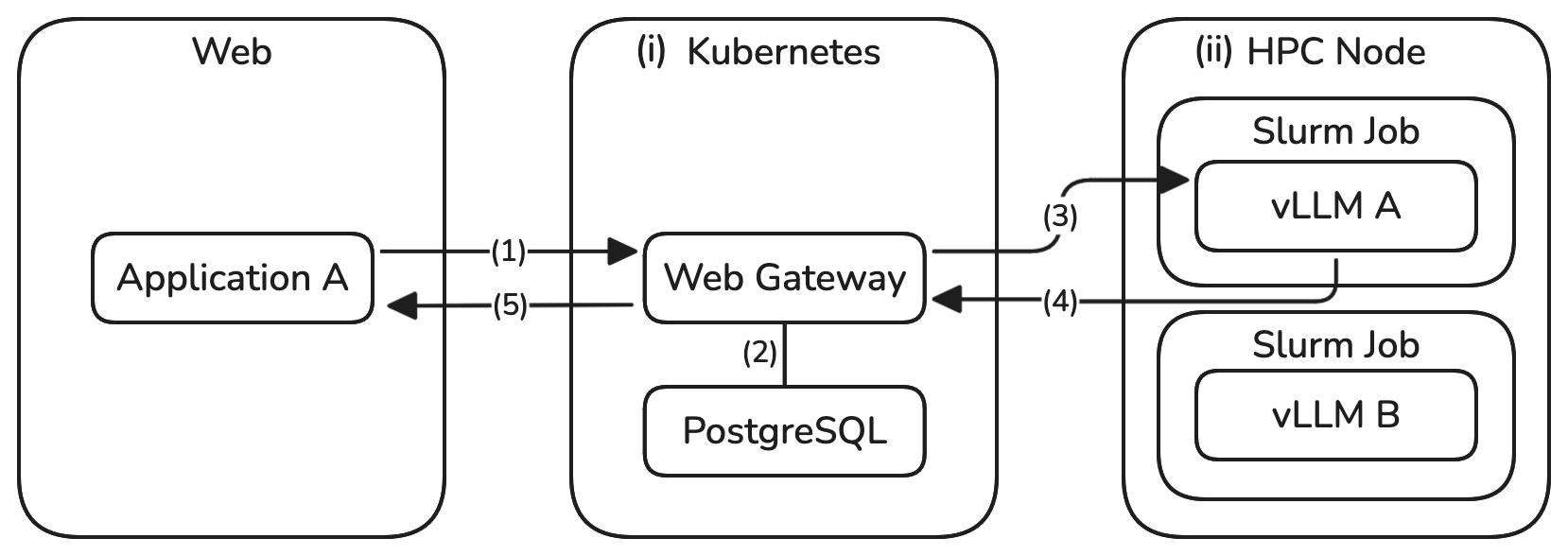}
  \Description{Describe the image}
  \caption{Visualization of the flow between layers for an incoming inference request.}
  \label{fig:webflow}
\end{figure}

Our approach separates various microservices into two layers, as illustrated in \Cref{fig:webflow}. The first layer consists of \textit{(i)} a Kubernetes-based web API that forwards inference requests from users to the corresponding models in the backend and manages the lifecycle of resources / models. The second layer consists of \textit{(ii)} the \ac{LLM} models themselves which are encapsulated into Slurm jobs.

\par To orchestrate the containers required for the services of our web API, we use Kubernetes\footnote{\url{https://kubernetes.io}} to automatically schedule, manage and scale the deployments. For the underlying container runtime we use Apptainer\footnote{\url{https://apptainer.org}}, since it is specifically designed for \ac{HPC} environments. As resource manager for the allocation and deployment of the \ac{HPC} nodes running the \ac{LLM} endpoints, we use Slurm\footnote{\url{https://slurm.schedmd.com}}, as this is the de facto standard in this domain.
 
\par
\begin{figure}[ht]
  \includegraphics[width=0.7\columnwidth]{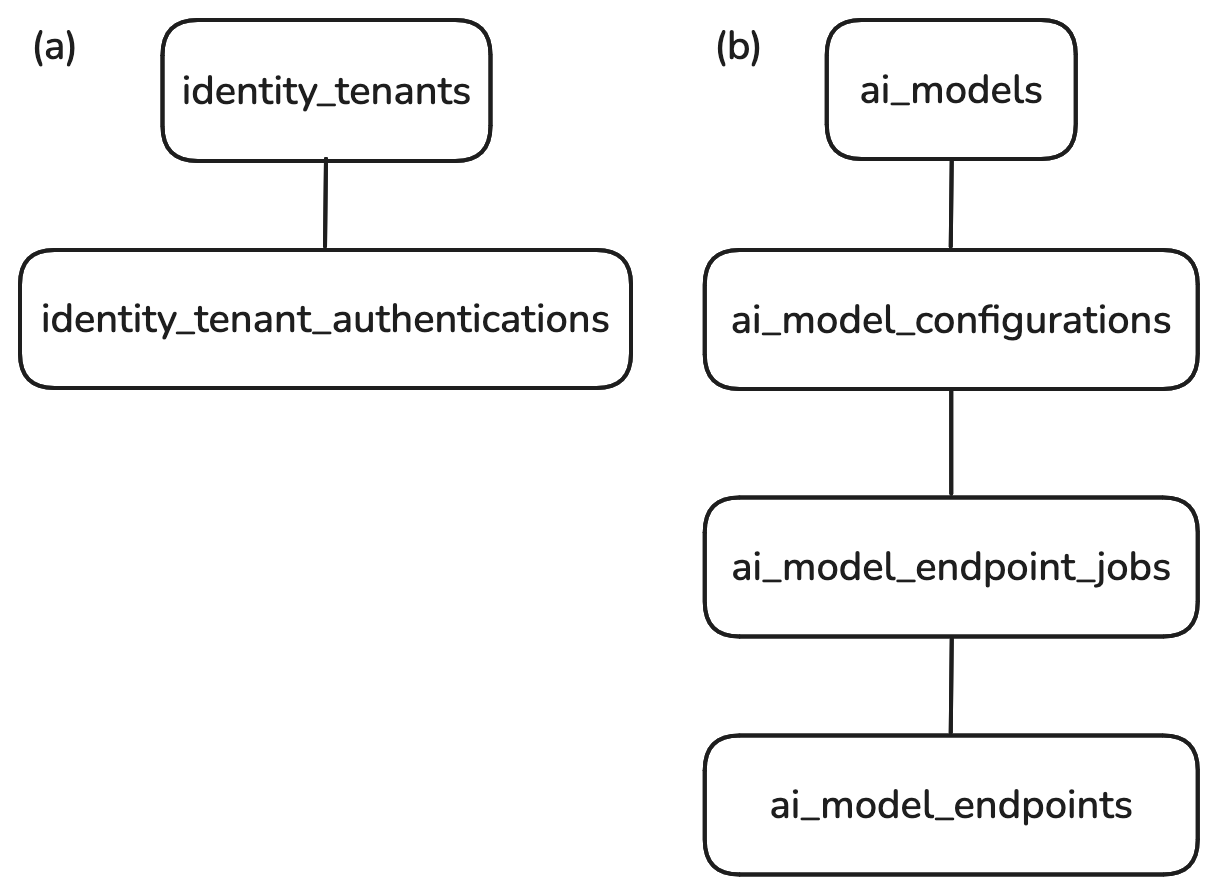}
  \Description{Describe the image}
  \caption{Visualization of the central database schema.}
  \label{fig:dbmodel}
\end{figure}

 Central to almost all of the microservices is a single relational PostgreSQL\footnote{\url{https://www.postgresql.org}} database running in Kubernetes. It holds two domains (): \textit{(a)} authentication and \textit{(b)} Slurm job management, as visualized in \Cref{fig:dbmodel}. Authentication consists of a simple 1:N relation of \texttt{identity\_tenants} and their API keys \texttt{identity\_tenant\_ authentications}, which are stored in an encrypted format. Slurm job management consists of multiple 1:N relations, which have proven to provide consistency in an actual production scenario. 

\subsection{Inference Components}
\subsubsection{vLLM}
Offering heterogeneous \ac{LLM} serving and inference in a high-through\-put scenario requires a fully scalable inference engine, such as the open-source software vLLM\footnote{\url{https://github.com/vllm-project/vllm}}. 
One of its major advantages is the introduction of \emph{PagedAttention}, which splits the \ac{KVC} into blocks and assigns them to logical pages allowing for dynamic scaling of allocated GPU memory \cite{kwon2023efficient}. 
This mapping is handled via block tables, managed by a central KV cache manager. In doing so, blocks may be reused in single- and shared across multiple requests, leading to significant performance gains in scenarios where single prompts consist of long inputs, or a large amount of requests need to be served simultaneously. This also allows for models with weight sizes exceeding the memory capacity of a single \ac{GPU} to run efficiently across multiple \acp{GPU}.  
vLLM implements an OpenAI-standard compatible FastAPI frontend for serving requests. If the number of requests received exceeds the system's concurrent throughput capabilities, a first-come, first-served scheduling policy is employed for all incoming requests. 

\subsubsection{Web Gateway}\label{ssec:webgateway}
The flow of a single request from a user's application through the Web Gateway to the desired vLLM endpoint running inside a Slurm job is visualized in \Cref{fig:webflow}.
The system's primary entry point for client applications is the Web Gateway, which is responsible for authentication and the subsequent forwarding of incoming requests. Its API endpoints are OpenAI-standard compatible, making them agnostic to the consuming client application. Request properties are strongly typed and validated, adding an additional layer of robustness. Authentication is handled via long-lived bearer-tokens, stored in an encrypted format in the database and authenticated on each request. To minimise database load, a distributed memory cache stores recently authenticated API keys for a limited period.
Subsequent to the authentication and validation process (1), the Web Gateway conducts a search for available endpoints corresponding to the requested LLM within the \texttt{ai\_model\_endpoints} table (2). It then forwards the request - along with all request parameters - to a matching node (3). The response flows from vLLM back to the Web Gateway (4) and then to the requesting application (5).
If no matching vLLM endpoint ready for inference is found, custom HTTP status codes are returned. 

\subsection{Management Components}
\begin{figure}[h]
  \includegraphics[width=0.9\columnwidth]{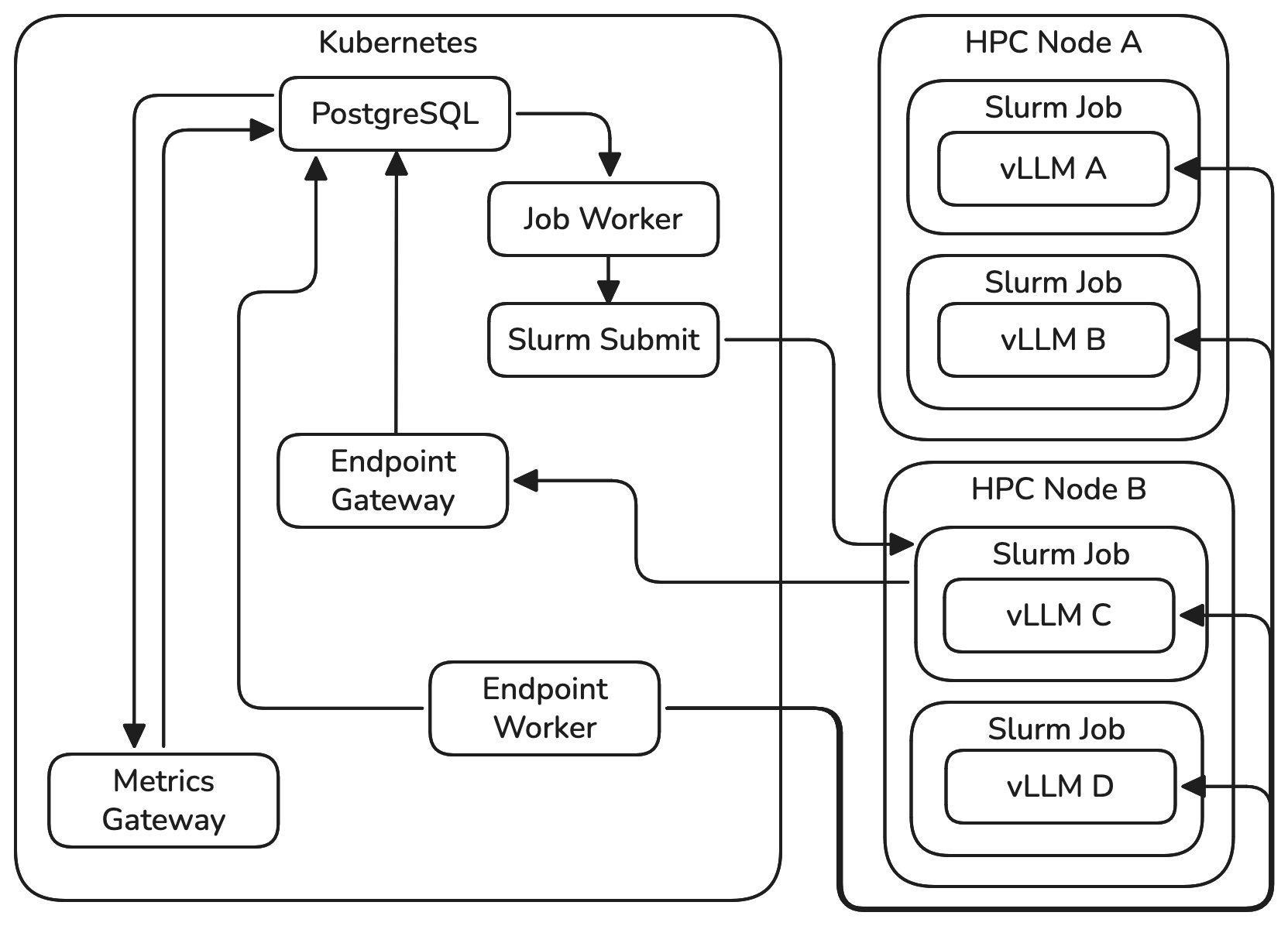}
  \Description{Describe the image}
  \caption{Visualization of the flow between layer for Slurm and vLLM management.}
  \label{fig:backendflow}
\end{figure}

 \Cref{fig:backendflow} visualizes the flow between microservices and layers described in the following subsections, which are ordered in the way the components interact with each other, starting from the Job Worker.

\subsubsection{Job Worker}\label{ssec:jobworker}
The Job Worker functions as an intermediary between our Kubernetes-based microservices and the Slurm-managed \ac{HPC} infrastructure. It is implemented as a long-running background process, responsible for scheduling new vLLM instances via Slurm and removing expired jobs.
Each run, it compares the entries in the \texttt{ai\_model\_endpoint\_jobs} table with the configurations specified in the \texttt{ai\_model\_configurations} table, for example number of requested model instances. 
If no matching endpoint job is found, configuration parameters are passed to the \nameref{ssec:slurmsubmit}, which requests ressourcess for a job with the corresponding model. On success, an \texttt{ai\_model\_endpoint\_jobs} entry is created, containing information such as the Slurm job id, timestamps for job submission, job registration and readiness of the contained vLLM server. 
To prevent simultaneous submission and subsequent startup of jobs, which could lead to inconsistent port mappings on compute-nodes, model configurations are iterated synchronously. The Job Worker waits for a specified timespan after a successful submit before proceeding to the next configuration. 

\subsubsection{Slurm Submit}\label{ssec:slurmsubmit}
Slurm Submit is the service handling communication between the \nameref{ssec:jobworker} and Slurm. It accepts a string from an SSH connection, which is forwarded to a bash script that actually deploys the job. A dedicated \texttt{munged} process inside the container provides authentication for the Slurm components. The comma-delimited parameters from the string are parsed and used to select a model-specific \texttt{.slurm} file from a folder mounted into the container, that in turn is used to run Slurm's \texttt{sbatch}. The file holds \verb+#SBATCH+ directives specifying hardware requirements of the requested node(s), among others. In addition, it sets environment variables for the vLLM container and configures a curl request to the \nameref{ssec:endpointgateway}. For multi-node multi-GPU setups, required for very large models, it also contains logic for setting up head and worker nodes.

\subsubsection{Endpoint Gateway}\label{ssec:endpointgateway}
Upon successful submission of a Slurm job, a curl \verb+POST+ is initiated from the Slurm script to the API of the Endpoint Gateway. It contains the endpoint job id, Slurm job id, node id, model version, capabilities and bearer token for the vLLM server that is being spawned. 
Authentication with the Endpoint Gateway is facilitated by a bearer token that is passed to the script. The underlying business logic proceeds to check if the internal job id holds a corresponding job in \texttt{ai\_model\_endpoint\_jobs} that has no endpoint attached. If that is the case, it compares the ports of all already existing endpoints on the supplied node and assigns a port \(p = argmax(port)+1\), which is passed back to the Slurm script in the curl response to be used by the vLLM server.
Finally, a new entry is created in \texttt{ai\_model\_endpoints}, holding information such as node, port, model version, bearer token and a nullable datetime field indicating readiness of the endpoint, which at that time is null. 

\subsubsection{Endpoint Worker}\label{ssec:endpointworker}
The Endpoint Worker is another long-running background service, responsible for managing endpoint health status. Each run, it iterates over all entries in the \texttt{ai\_model \_endpoint\_jobs} table and sends a \verb+GET+ request to the \verb+/health+ endpoint of each job, which is provided by the vLLM server. 
Endpoint jobs not yet marked as ready but receive a response status-code \verb+200+ are updated accordingly by setting their corresponding fields to the current datetime. Once marked ready, the endpoint will be considered for forwarding requests by the \nameref{ssec:webgateway}.
In cases where no response is returned, the Endpoint Worker differentiates two cases: (1) jobs that have been canceled or expired and (2) jobs that are still starting up.
Since loading model weights can require a substantial amount of time, a configurable 30-minute timeout is currently implemented before a job is deemed cancelled or expired. Moving forward, this could be dynamically configured with an additional estimated load time on a per-model basis via the \texttt{ai\_model\_configurations} table.
For cancelled or expired jobs, the Endpoint Worker removes the corresponding entries in \texttt{ai\_model\_endpoints} and \texttt{ai\_model\_endpoint\_jobs}. 

\subsubsection{Metrics Gateway}\label{ssec:metricsgateway}
The Metrics Gateway serves API endpoints for Prometheus, a time-series database, and Grafana, an open-source observability platform.
The Prometheus endpoint returns a response with the required format for Prometheus' HTTP service discovery, which allows for dynamic (de-)registration of targets for metrics scraping. By accessing the \texttt{ai\_model\_endpoints} table, it converts necessary information like node id, port and bearer token of all running vLLM instances to the specified Prometheus template and allows for additional meta fields to be added, such as job id and Slurm job id. While metrics of our Kubernetes based microservices can be conveniently scraped with Prometheus Operator, obtaining metrics from the vLLM instances requires this workaround, as they are not part of the Kubernetes cluster and may change network adresses over time.
The Grafana endpoints, in contrast, accept various \verb+POST+ request parameters, which are used to control the amount and type of running model instances.

\subsection{Observability \& Automated Dynamic Scaling}\label{sec:observabilityscaling}
In addition to the fundamental inference and management components, an observability stack is utilized, comprising Prometheus, Grafana and Grafana Loki to facilitate visualisation, reporting and logging. This allows for real-time monitoring of critical metrics, such as the vLLM instance load or concurrent \nameref{ssec:webgateway} requests, hence providing insight into how frequently certain models are accessed. In combination with live logging, it provides the capability to efficiently identify and resolve issues almost instantaneously.
\par This stack also comprises the central part of the mechanism to dynamically scale up or down vLLM model instances inside Slurm jobs, based on GPU load. Utilizing Grafanas alert rules with its contact point webhook notifications, custom JSON payloads are automatically sent to the metrics gateway. Here, the business logic adjusts the number of model instances specified in the \texttt{ai\_model\_job \_configurations} table. This triggers the \nameref{ssec:jobworker} to automatically start the newly requested number of instances on its next invocation, currently every 15 seconds.
Hence, the system scales by actual hardware load, since vLLM reports metrics like KVC utilization, queue times and token throughput. In our first tests, we utilized the vLLM queue time metric, where a queue time above 5 seconds over 30 sustained seconds triggered instantiation of an additional model instance (see \autoref*{ssec:results} for further explanation). Compared to scaling by number of requests, which in the context of \acp{LLM} is an arbitrary metric on its own given the lack of uniformity of the requests \cite{BurstGPT}, for example input token count or hyperparameters like maximum generated tokens, this setup allows for maximizing GPU load and thus token throughput. In our first tests, we utilized the vLLM queue time metric, where a queue time above 5 seconds over 30 sustained seconds triggers instantiation of an additional model instance (c.f. \autoref*{ssec:results}).

%% file: content/04_preliminary_results.tex
\section{Performance Evaluation}
\label{sec:evaluation}
\subsection{Methodology}
In this section, we provide preliminary performance results for our system via the serve-benchmark in vLLM using the \emph{BurstGPT\_without\_fails\_2} dataset \cite{BurstGPT}.

We benchmark two system configurations: (1) \emph{GPU-S}, which utilizes two NVIDIA L40S GPUs with an AMD EPYC 9654 CPU, and (2) \emph{GPU-L}, which uses one NVIDIA H100 GPU and an AMD EPYC 9454 CPU. All configurations utilize 96GB RAM. Networking is based on Mellanox Infiniband HDR100 (100GBit/s), with two ports for redundancy on Kubernetes nodes and one port on compute nodes. The topology consists of 8 leaf-switches linked to 2 spine-switches via 5 HDR (200GBit/s) links each, resulting in a bisection bandwidth of 8000 Gb/s. The maximum blocking factor is 2.3:1.

\par For our baseline model, we use Mistral Small 3.2 24B Instruct 2506\footnote{\href{https://huggingface.co/mistralai/Mistral-Small-3.2-24B-Instruct-2506}{Mistral Small 3.2 24B Instruct 2506 on HuggingFace}} with default parameters. The load is simulated and averaged over 50 runs for three scenarios: 100, 500 and 1000 concurrent requests.
All runs are performed from a dedicated interactive node inside the HPC cluster, using the same vLLM container image running inside the Slurm jobs used for inference serving (v0.10.2). 
The seed is set to 0, to ensure that every benchmark run uses the same samples from the dataset. The vLLM node is not restarted or modified between runs; however, between each run, the KVC is fully cleared. All requests are handled using streaming.
\par In addition to the baseline benchmark, the system is also benchmarked including the Web Gateway. Before starting a full run, the vLLM serve-benchmark sends one initial request to the specified target, which triggers the caching of the authentication in the Web Gateway, causing it to only perform one database trip for each incoming request to look up the vLLM node endpoint. 

\input{tables/large_table}
\subsection{Results}\label{ssec:results}
Results for the GPU-S and GPU-L configurations are shown in \autoref{table:benchmarks_gpu_all}. The vLLM benchmark suite defines time to first token (TTFT) as a timespan between sending the request and receiving the corresponding first token. 
Similarly, end to end latency (E2EL) is a timespan between sending the request and receiving the corresponding last token.
Therefore time per output token (TPOT) is defined as:
\begin{equation}
\mathtt{tpot} = \frac{\mathtt{e2el} - \mathtt{ttft}}{\mathtt{output\_len} - 1}
\end{equation}

Unintuitively, for GPU-S, total request duration and TPOT favors the Web Gateway setup, albeit with slightly increased E2EL in 100 and 1000 concurrent request scenarios. The 500 concurrent request scenario stands out particularly for TTFT, where sending requests to vLLM directly is noticeably faster by about 500ms ($23.34\%$). GPU-L, in contrast, behaves more as expected, displaying slightly higher total request durations across all Web Gateway scenarios, with a higher E2EL and TTFT from 500 concurrent requests upward. Curiously, TPOT is significantly lower in these cases ($37.21\%$ / $43.45\%$). The reasons for this will need to be investigated in a more detailed system analysis in the future. 
Inspecting the vLLM logs, the GPU-S configuration starts queuing requests from 255 concurrent requests onward, while the GPU-L configuration does not appear to queue requests in any of our scenarios. While this could be an indicator for queueing impacting general performance, especially when done across multiple GPUs, it also leads to the assumption that once queueing has started, the slight latency introduced by the Web Gateway could buffer the impact of load put on the vLLM api server, `masking' the queuing process. When compared to GPU-L (where no queueing occurs), we observe higher E2EL and TTFT as expected. However, at 500 and 1000 concurrent requests, MD and Std for TTFT deviates massively for the Web Gateway when compared to direct vLLM node access, with a nearly 1 second ($41.95\%$) slower median TTFT.

This suggests that our current implementation of routing requests via the Web Gateway starts to bottleneck in scenarios where enough GPU compute is present to handle the incoming load. 

%% file: tables/large_table.tex
\begin{table*}[t!]
\caption{GPU-S and GPU-L configuration performance benchmarks for concurrent vLLM and Web Gateway requests.}
\scriptsize
\centering
\begin{tabular}{l|llllll|llllll}
\toprule
Configuration & \multicolumn{6}{c}{GPU-S} &  \multicolumn{6}{c}{GPU-L} \\
Benchmark & \multicolumn{3}{c}{vLLM Node} & \multicolumn{3}{c}{Web Gateway} & \multicolumn{3}{c}{vLLM Node} & \multicolumn{3}{c}{Web Gateway} \\
Concurrent Requests & 100 & 500 & 1000 & 100 & 500 & 1000 & 100 & 500 & 1000 & 100 & 500 & 1000 \\
\midrule
E2EL Median (ms) & 2307.96 & 7381.58 & 22385.04 & 2186.39 & 6331.33 & 22963.34 & 1195.80 & 3029.86 & 6190.10 & 1149.93 & 3076.11 & 9419.23 \\
E2EL Std (ms) & 5215.32 & 9316.33 & 20958.25 & 5151.45 & 9148.52 & 19999.29 & 2940.16 & 4984.30 & 8206.37 & 2981.23 & 4994.50 & 8465.61 \\
Requests Total Duration (s) & 30.01 & 48.02 & 84.76 & 30.03 & 46.95 & 79.75 & 17.20 & 25.94 & 34.81 & 17.31 & 26.28 & 37.25 \\
Requests Total Input Tokens & 77561.00 & 381456.00 & 768960.00 & 77561.00 & 381456.00 & 768960.00 & 77561.00 & 381456.00 & 768960.00 & 77561.00 & 381456.00 & 768960.00 \\
Requests Total Output Tokens & 7048.98 & 49763.68 & 141407.84 & 6976.34 & 49567.64 & 141313.34 & 6978.78 & 50126.12 & 142778.08 & 7027.58 & 49955.58 & 142664.64 \\
TPOT Median (ms) & 65.52 & 90.61 & 102.13 & 63.33 & 79.20 & 94.97 & 33.41 & 67.21 & 86.74 & 32.49 & 42.20 & 49.05 \\
TPOT Std (ms) & 6.36 & 13.07 & 18.00 & 5.52 & 6.92 & 10.50 & 2.34 & 12.19 & 29.23 & 2.08 & 3.16 & 5.58 \\
TTFT Median (ms) & 404.38 & 1849.79 & 8328.36 & 334.79 & 2412.93 & 6825.14 & 225.58 & 993.91 & 2190.97 & 207.35 & 1133.02 & 3109.98 \\
TTFT Std (ms) & 37.59 & 1477.46 & 12460.81 & 146.29 & 1950.86 & 13191.81 & 14.34 & 87.98 & 189.22 & 62.97 & 841.35 & 3642.25 \\
Throughput Requests (req/s) & 3.33 & 10.54 & 11.83 & 3.33 & 10.77 & 12.56 & 5.82 & 19.52 & 28.87 & 5.79 & 19.31 & 26.95 \\
Throughput Token Output (tok/s) & 234.91 & 1048.13 & 1672.62 & 232.51 & 1066.95 & 1775.28 & 405.79 & 1956.22 & 4121.70 & 406.30 & 1928.61 & 3844.97 \\
Throughput Token Total (tok/s) & 2820.34 & 9088.70 & 10768.98 & 2818.61 & 9282.65 & 11435.67 & 4916.07 & 16849.58 & 26322.19 & 4893.63 & 16664.20 & 24569.65 \\
\bottomrule
\end{tabular}
\label{table:benchmarks_gpu_all}
\end{table*}

%% file: content/04_discussion.tex
\section{Discussion}
\label{sec:discussion}
The presented architecture is still considered a work in progress, and this paper does not claim to represent best practices. Its purpose is to provide a starting point for other infrastructure projects, highlighting where the chosen implementation has succeeded, where it faced problems, and where it currently falls short. 

From early friendly user tests with small groups that use various applications to access our \nameref{ssec:webgateway}, such as KI:Connect\footnote{\url{https://kiconnect.pages.rwth-aachen.de/pages/}} or Open WebUI\footnote{\url{https://openwebui.com}}, we are confident that the foundation of our architecture - scaling LLM serving for synchronous applications in an HPC environment - is sound. Management of vLLM instances in Slurm jobs has run successfully in this small-scale production setting for multiple months, across a variety of tested models, covering single-node single-GPU, single-node multi-GPU, and even multi-node multi-GPU configurations. 

Our benchmarks show that when scaling beyond a 1000 concurrent request scenario, improvements must be made to our benchmark setup and/or the \nameref{ssec:webgateway}. The increase in TTFT degrades end-user experience, signaling a longer wait time between the user sending a prompt and the model starting to return a response. However, TPOT remains consistent, which translates to smooth streaming and therefore `typing'-animation for chatbot applications. 
We identify several key areas that require further investigation to solve the arising bottleneck:  

\textbf{Networking}: since we run the vLLM benchmark suite from within a compute-node of the HPC system, high concurrency scenarios might be impacted by networking components, favoring a direct request from compute-node to compute-node. Alternatively, improper request handling in the \nameref{ssec:webgateway} may contribute.  

\textbf{Caching}: as every request performs a database lookup for a matching model endpoint, implementing a caching mechanism could reduce database load and speed up request handling.  

\textbf{Scaling}: horizontal or vertical scaling of either the \nameref{ssec:webgateway} or PostgreSQL via Kubernetes could improve performance.  

Despite these necessary improvements, institutional provisioning of \acp{LLM} on sovereign infrastructure provides clear data privacy benefits, especially when handling sensitive data. However, preventing \ac{LLM}-based attack vectors like prompt injections requires trained experts and domain specific knowledge, which must be seen as additional cost for said privacy benefits.

%% file: content/05_conclusion.tex
\section{Conclusion \& Outlook}
\label{sec:outlook}
In this work, we presented our approach to building a production-ready, modular architecture for highly scalable serving of LLM inference in an HPC context. By leveraging HPC-native components like Slurm and Apptainer, managed by custom tailored microservices and integrating them with cloud-native solutions for orchestration and observability like Kubernetes, Prometheus and Grafana, we implemented a robust solution to deploy and scale LLMs using the open-source inference engine vLLM. 
In addition, we shared performance benchmarks for two distinct compute-node hardware configurations under different load scenarios, showcasing the viability of HPC infrastructure for synchronous workloads like web requests from AI chatbot applications. 
As our architecture represents a work in progress, our next steps will be directed toward a high-troughput production scenario in a higher education setting. Given the scarcity of usage data for this application, we hope to attain a better understanding of how far sovereign infrastructure for AI applications needs to scale. We will also explore the deployment of other generative ai modalities.
With growing concerns about the sustainability of AI, ensuring efficient usage of compute resources will take high priority. Thus, further work on leveraging components like Slurm to balance compute during peak usage times of AI inference while allocating the unused ressources to research computations during off-hours should take priority. Furthermore, exploring hardware configurations from vendors other than NVIDIA seems advisable to prevent a vendor lock-in for datacenters.

%% file: content/acknowledgements.tex
\begin{acks}
  The work presented in this paper is the result of the joint project by Ruhr-University Bochum and University of Cologne: `Open Source-KI.nrw'. The project is fully funded by the Ministry of Culture and Science of the State of North Rhine-Westphalia. 
\end{acks}